\documentclass[11pt,a4paper]{article}
\usepackage[T1]{fontenc}
\usepackage[utf8]{inputenc}
\usepackage[lmargin=1in,rmargin=1in,tmargin=1in,bmargin=1in]{geometry}
\usepackage[english]{babel}
\usepackage{graphicx}
\usepackage{tabularx}

\usepackage{cellspace}
\addparagraphcolumntypes{X}
\usepackage{caption}
\usepackage{prettyref}
\newrefformat{fig}{Fig.~\ref{#1}}
\usepackage{xcolor}
\usepackage{mathtools}

\DeclareMathOperator{\ch}{ch}
\DeclareMathOperator{\Log}{Log}
\DeclareMathOperator{\tth}{th} 
\DeclareMathOperator{\tg}{tg}

\usepackage[noorphans,font=itshape,indentfirst=false]{quoting}

\newcommand{\opT}{\mathrm{T}} 

\newcommand*\diff{\mathop{}\!\mathrm{d}} 

\usepackage[super,comma,sort&compress]{natbib}

\usepackage[colorlinks,allcolors=blue]{hyperref}

\usepackage{cleveref}

\title{Cancellation of the central singularity of the Schwarzschild solution with natural mass inversion process.\\{\small \textit{\textbf{Published in}Modern Physics Letters A, Vol. 30, No. 9 (2015) 1550051}}}
\author{J.P. Petit\thanks{CNRS, BP 55, 84122 Pertuis, France, e-mail: \href{mailto:jppetit1937@yahoo.fr}{jppetit1937@yahoo.fr}} \and G. D'Agostini\thanks{dagostini.gilles@laposte.net}}
\date{\today}

\begin{document}
\maketitle

\begin{abstract}
	We reconsider the classical Schwarzschild solution in the context of a Janus cosmological model. We show that the central singularity can be eliminated through a simple coordinate change and that the subsequent transit from one fold to the other is accompanied by mass inversion. In such scenario matter swallowed by black holes could be ejected as invisible negative mass and dispersed in space.
\end{abstract}

{\small\emph{Keywords:} Black hole; space bridge; Schwarzschild metric; Kerr metric; central singularity; Janus cosmological model; Gaussian coordinates; mass inversion process.}

PACS Nos.: 98.80.Bp, 98.80.Qc

\section{Introduction}

When Schwarzschild published his solution\cite{schwarzschild} of Einstein equation, in 1916, the basic hypothesis was just time-independence and spherical symmetry. Nobody suspected that there was an additional one: null-homotopy. The consequence of this last one was that, in the proposed solution, $r$ was a radial coordinate:

\begin{equation}
	\diff s^2=\left(1-\frac{R_{\mathrm{s}}}{r}\right)c^2\diff t^2-\frac{\diff r^2}{\left(1-\frac{R_{\mathrm{s}}}{r}\right)}-r^2\left(\diff\theta^2+\sin^2\theta\diff\varphi^2\right).
\end{equation}

A first problem at $r = R_{\mathrm{s}}$, considered as coordinate singularity, was eliminated by various coordinates changes. See \prettyref{fig:table}.

\begin{figure}
	\begin{tabularx}{\linewidth}{|>{\centering}m{3.75cm}|Sc|S{X}|}
	\hline
	Coordinates & Line element & \ \\
	\hline
	Eddington-Finkelstein (ingoing) & $\displaystyle\left(1-\frac{r_{\mathrm{s}}}{s}\right)\diff v^2-2\diff v \diff r -r^2\diff\Omega^2$ & regular at horizon, extends across future horizon \\
	\hline
	Eddington-Finkelstein (outgoing) & $\displaystyle\left(1-\frac{r_{\mathrm{s}}}{s}\right)\diff u^2-2\diff u \diff r -r^2\diff\Omega^2$ & regular at horizon, extends across \mbox{past horizon} \\
	\hline
	Gullstrand-Painlevé & $\displaystyle\left(1-\frac{r_{\mathrm{s}}}{s}\right)\diff T^2-2\sqrt{\frac{\tau_{\mathrm{s}}}{r}}\diff T\diff r -\diff r^2-r^2\diff\Omega^2$ & regular at horizon \\
	\hline
	Isotropic & $\displaystyle\frac{\left(1-\frac{r_\mathrm{s}}{4R}\right)}{\left(1+\frac{\tau_\mathrm{s}}{4R}\right)^2}\diff t^2-\left(1+\frac{r_\mathrm{s}}{4R}\right)^4\left(\diff x^2+\diff y^2+\diff z^2\right)$ & isotropic lightcones on constant \mbox{time slices} \\
	\hline
	\end{tabularx}
	
	\caption{\label{fig:table}Schwarzschild’s geometry. Alternative coordinates.}
\end{figure}

But $r = 0$ was considered as a \emph{true singularity}, a point of space where the geodesics end. The goal of this paper is to show, through a very simple coordinate change, that this so-called ``central singularity'' is due to a wrong choice of local topology.

We must introduce the concept of \emph{representation space} which is the space in which we build a mental image of an object. As an example we believe we live in R3 and that space and time are separated. This old belief is useful and enough for today's life. But if the speed of light was much smaller we should shift to Minkowski's spacetime. Similarly, when we send a probe to Jupiter or predict an eclipse we neglect space curvature. If this last was not neglectible, we should use Einstein's equation instead of Newton's. If we intend to describe an object through a line element it is easy to show, through 2D examples that inadequate coordinate choice may induce wrong image of a geometric object. For example, consider the following metric:
\begin{equation}\label{eq2}
	\diff s^2=\frac{\diff r^2}{1-\frac{r^2}{R_{\mathrm{s}}^2}}+r^2\diff\varphi,
\end{equation}
whose signature $(+, +)$ changes into $(-, +)$ when $r > R_{\mathrm{s}}$. But, through the simple following coordinate change
\begin{equation}
	r=R_{\mathrm{s}}\sin\theta,
\end{equation}
this object becomes a sphere:
\begin{equation}
	\diff s^2=R_{\mathrm{s}}^2\left(\diff\theta^2+\sin^2\theta\diff\varphi^2\right).
\end{equation}

Another example. Consider the metric:

\begin{equation}
	\diff s^2=\frac{\diff r^2}{-r^2+2rR+r_0^2-R^2}+r^2\diff\varphi^2
\end{equation}

The signature is $(+, +)$ if
\begin{equation}
	R-r_0<r<R+r_0,
\end{equation}
but through the new coordinate change
\begin{equation}
	r=R+r_0 \cos\theta,
\end{equation}
we get the well-known metric of the torus:
\begin{equation}
	\diff s^2=r_0 \diff\theta^2+\left(R+r_0 \cos\theta\right)\diff\varphi^2.
\end{equation}

Now, let us focus on the space part of Schwarzschild's line element, limited to $\{r, \varphi\}$ coordinates:
\begin{equation}
	\diff\Sigma^2=\frac{\diff r^2}{1-\frac{R_{\mathrm{s}}}{r}}+r^2\diff\varphi^2
\end{equation}

For $r < R_{\mathrm{s}}$, the signature $(+, +)$ is changed into $(-, +)$. Let us make the change of variable:
\begin{equation}\label{eq10}
	r=R_{\mathrm{s}}\left(1+\Log\ch\rho\right),
\end{equation}
which gives:
\begin{equation}
	\diff\Sigma^2=R_{\mathrm{s}}^2\left[\frac{\left(1+\Log\ch\rho\right)}{\Log\ch\rho}\tth^2\rho\diff\rho^2+\left(1+\Log\ch\rho\right)^2\diff\varphi^2\right].
\end{equation}

All singularities disappear, $r = R_{\mathrm{s}}$ corresponds to $\rho = 0$. In this point the determinant of the metric
\begin{equation*}
	\det g=R_{\mathrm{s}}^4\frac{\left(1+\Log\ch\rho\right)^2}{\Log\ch\rho}\tth^2\rho,
\end{equation*}
is no longer zero. The metric is well defined for all values of $\rho$. If we embed the surface in a 3D-Euclidean space we can define the meridians, corresponding to
\begin{equation}
	\diff\Sigma^2=\frac{\diff r^2}{1-\frac{R_{\mathrm{s}}}{r}}+\diff z^2
\end{equation}

And we immediately get the meridians as\footnote{Erratum page~\pageref{erratum}}
\begin{equation}\label{eq13}
	z=\pm2R_{\mathrm{s}}\sqrt{\frac{r}{R_{\mathrm{s}}}-1},\quad r^2=R_{\mathrm{s}}+\frac{z^2}{4R_{\mathrm{s}}}.
\end{equation}

The surface is a \emph{space bridge}, a ``2D diabolo'' linking two 2D-Euclidean surfaces. 

\begin{figure}[hb]
	\begin{center}
		\includegraphics[width=7cm]{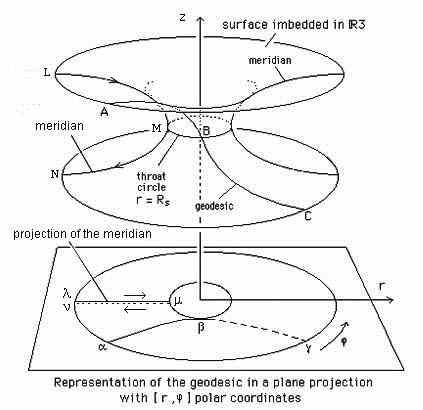}
	\end{center}
	\caption{\label{fig:diabolo} The 2D diabolo embedded in $R^3$.}
\end{figure}

The problem of the signature has disappeared. From Lagrange equations we can calculate the geodesics in the $[\rho, \varphi]$ coordinate system. If embedded, the surface owns a throat circle whose perimeter is $2\pi R_{\mathrm{s}}$. We can shape the surface as a twofold $F^{(+)}$ and $F^{(-)}$ cover of a $M_2$ manifold with a 1D common circular border, and create induced mapping between adjacent points $M^{(+)}$ and $M^{(-)}$.

\begin{figure}[t]
	\begin{center}
		\includegraphics[width=10cm]{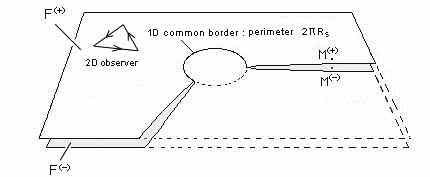}
	\end{center}
	\caption{\label{fig:twofold} Twofold cover of a manifold with a circle as common boundary.}
\end{figure}

\begin{figure}[t]
	\begin{center}
		\includegraphics[width=10cm]{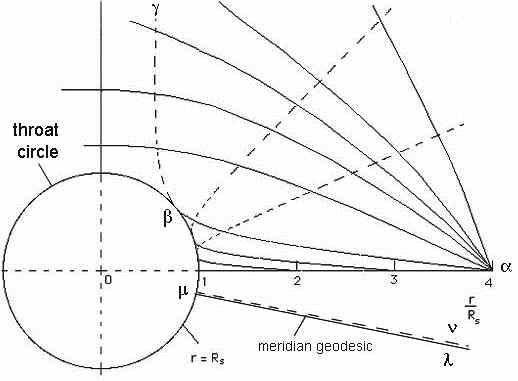}
	\end{center}
	\caption{\label{fig:projection} Projection of geodesics.}
\end{figure}

Figure \ref{fig:projection} shows the projection $\alpha\beta\gamma$ of a geodesic $\mathrm{ABC}$ (see \prettyref{fig:diabolo}) tangent to the common boundary, which ensures the continuity of the geodesics of the fold $F^{(+)}$ with the ones of the fold $F^{(-)}$ (dotted lines). Radial line $\lambda \mu \nu$ is the image of a geodesic meridian curve $\mathrm{L M N}$ of \prettyref{fig:diabolo}. This will be important for the following.

\begin{figure}
	\begin{center}
		\includegraphics[width=8cm]{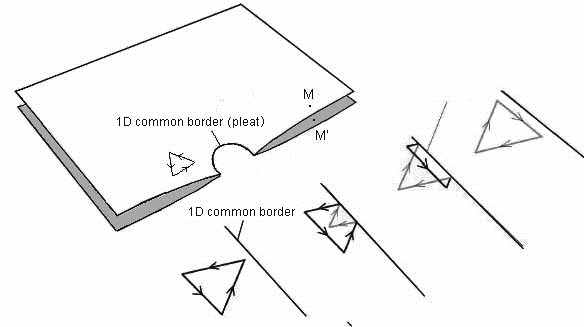}
	\end{center}
\caption{\label{fig:triangle}When the triangle crosses the common boundary, the orientation is reversed.}
\end{figure}

Figure \ref{fig:triangle} shows how vicinities of adjacent points are linked by enantiomorphy relationship.

\Cref{fig:twofold,fig:projection,fig:triangle} correspond to a reduction of our representation space to a 2D Euclidean representation space, through a non-isometric imbedding which artificially transforms the throat into a pleat and couples points through an adjacent and enantiomorphic relationship. ``The circle owns no center and there is nothing inside'', because we are out of the considered 2D surface, the 2D diabolo.

Now introduce the 3D metric
\begin{equation}
	\diff\Sigma^2=\frac{\diff r^2}{1-\frac{R_{\mathrm{s}}}{r}}+r^2\left(\diff\theta^2+\sin^2\theta\diff\varphi^2\right),
\end{equation}
which is Euclidean at infinity. When $r < R_{\mathrm{s}}$ the signature $(+ + +)$ is changed into $(- + +)$. Applying \eqref{eq10} we get
\begin{equation}
	\diff s^2=R_{\mathrm{s}}^2\left[\frac{\left(1+\Log\ch\rho\right)}{\Log\ch\rho}\tth^2\rho\diff\rho^2+\left(1+\Log\ch\rho\right)^2\left(\diff\theta^2+\sin^2\theta\diff\varphi^2\right)\right].	
\end{equation}

Its determinant never vanishes. The metric is well defined for all values of $\rho$ and is Euclidean at infinite. It is a \emph{3D space bridge} linking to 3D Euclidean spaces. Its throat, corresponding to $\rho = 0$, is a S2 sphere.

\begin{figure}
	\begin{center}
		\includegraphics[width=6cm]{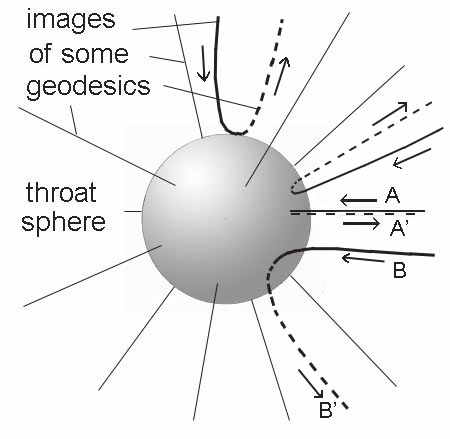}
	\end{center}
\caption{\label{fig:sphere} When geodesics of the 3D hypersurface cross the throat sphere they appear tangent to it in this 3D non-isometric imbedding in a 3D Euclidean space.}
\end{figure}

Unfortunately we do not own a 4D Euclidean representation space to operate an isometric imbedding. So that, similarly as in \prettyref{fig:twofold}, we will use a non-isometric imbedding in 3D Euclidean space. Then the throat is converted into a ``pleat'' along the projected S2 sphere. Similarly, some geodesics, continuous in this 3D hypersurface, seem to turn back on the sphere (see \prettyref{fig:sphere}).

\begin{figure}
	\begin{center}
		\includegraphics[width=5cm]{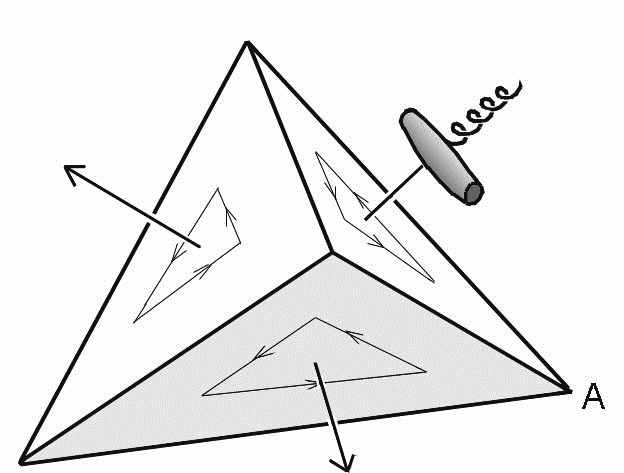}
	\end{center}
\caption{\label{fig:tetrahedron} Oriented tetrahedron.}
\end{figure}

\begin{figure}
	\begin{center}
		\includegraphics[width=12cm]{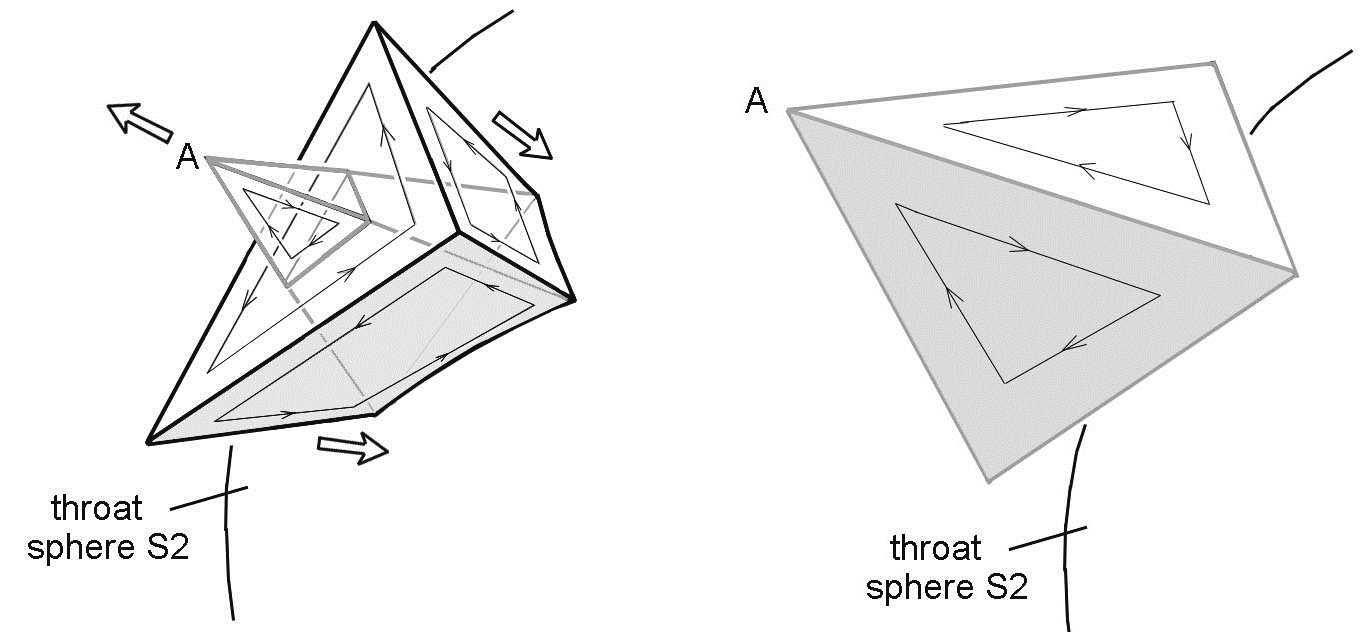}
	\end{center}
\caption{\label{fig:inverted} By crossing the throat sphere, the tetrahedron is inverted.}
\end{figure}

To show 3D space orientation we will use a tetrahedron (see \prettyref{fig:tetrahedron}). To illustrate the enantiomorphy relationship we need to project this object through the S2 throat sphere, each vertice following a geodesic, as showed in \prettyref{fig:inverted}.

In the 2D $\{\rho, \varphi\}$ representation the adjacent points are defined by the relation:
\begin{equation*}
	M\colon(\rho,\theta)\rightarrow M'\colon(-\rho,\theta).
\end{equation*}

In the 3D $\{\rho, \theta, \varphi\}$ representation the adjacent points in 2D and 3D are defined by the relation:
\begin{equation*}
	M\colon(\rho,\theta,\varphi)\rightarrow M'\colon(-\rho,\theta,\varphi).
\end{equation*}

The association of points $M$ and $M'$ goes hand in hand with an enantiomorphic relation between their corresponding neighborhoods.

Now let us go back to \eqref{eq2} and apply \eqref{eq10}. We get
\begin{equation}\label{eq16}
	\diff s^2=\frac{\Log\ch\rho}{1+\Log\ch\rho}c^2\diff t^2-R_{\mathrm{s}}^2\left[\frac{\left(1+\Log\ch\rho\right)}{\Log\ch\rho}\tth^2\rho\diff\rho^2+\left(1+\Log\ch\rho\right)^2\left(\diff\theta^2+\sin^2\theta\diff\varphi^2\right)\right].
\end{equation}

When $\rho$ tends to $\pm\infty$, $\Log\ch\rho \rightarrow \rho$ and $\tth\rho \rightarrow 1$. The metric tends to Lorentz
metric. Space is extended to $(\rho > 0; \rho < 0)$ domain. The hypersurface becomes a \emph{spacetime bridge}, linking two Lorentz spaces through a throat surface S2. When we calculate the geodesics in the plane $\theta = \frac{\pi}{2}$ in the $\{t, r, \theta, \varphi\}$ representation we find  the following (Eq. (6.90) in Ref. \citenum{abs}):
\begin{equation}
	\diff\varphi=\pm\frac{1}{r^2}\frac{\diff r}{\sqrt{\frac{c^2 l^2 -1}{h^2}+\frac{R_{\mathrm{s}}}{h^2 r}-\frac{1}{r^2}+\frac{R_{\mathrm{s}}}{f^3}}},
\end{equation}
where $l$ and $h$ are the classical parameters of the quasi-Keplerian trajectory (Eqs. (6.80) and (6.81) in Ref. \citenum{abs}). On the Schwarzschild's sphere $(r = R_{\mathrm{s}})$ we get:
\begin{equation}
	\tg\alpha=R_{\mathrm{s}}\left\lvert\frac{\diff\varphi}{\diff r}\right\rvert_{r=R_{\mathrm{s}}}=\frac{h}{R_{\mathrm{s}}cl}.
\end{equation}

In a $\{t, \rho, \theta, \varphi\}$ representation, with $\theta = \frac{\pi}{2}$, we obtain curves $\rho = f(\varphi)$ that will be 
inscribed in two (adjacent) folds $F^{(+)}$ and $F^{(-)}$.

With
\begin{equation}
	\left(\tg\beta\right)_{\rho\rightarrow 0}=\frac{\rho\diff\varphi}{\diff\rho}=\frac{r\diff\varphi}{\diff r}\frac{\rho}{r}\frac{\diff r}{\diff\rho}=\frac{h}{R_{\mathrm{s}}^2 cl}\rho^2\rightarrow0.
\end{equation}

\begin{figure}
	\begin{center}
		\includegraphics[width=7cm]{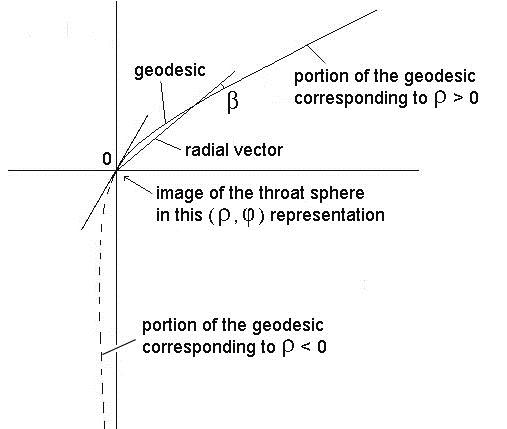}
	\end{center}
\caption{\label{fig:representation} Geodesical path in a $[\rho,\varphi]$ representation (plane, $\theta=\pi/2$).}
\end{figure}

The throat sphere $r = R_{\mathrm{s}}$ is reduced to a point. But the geodesics of the fold $F^{(+)}$ can be prolonged continuously in the adjacent fold $F^{(-)}$. \emph{The central singularity disappears}. Now let us deal with spacetime structures.

Referring to \emph{Introduction to General Relativity}, Sec. 2.6 of Ref. \citenum{abs}:

\begin{quote}

\begin{figure}[b]
	\begin{center}
		\includegraphics[width=8cm]{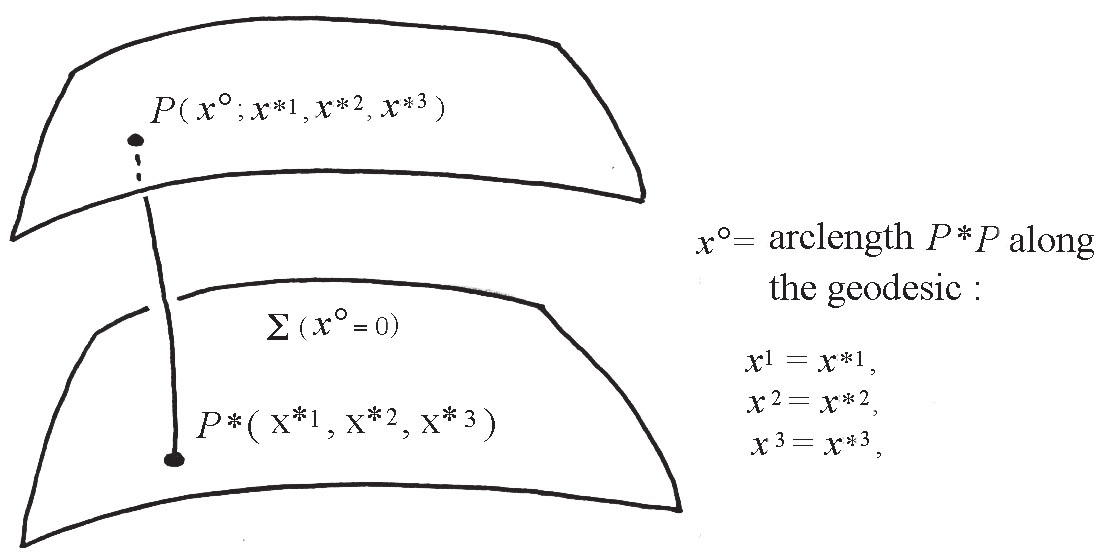}
	\end{center}
\caption{\label{fig:arclenght}}
\end{figure}

By letting a family of geodesics play a particular role among the coordi-nates lines Gauss introduced a useful coordinate system. Consider a 4D space with hyperbolic metric with signature $(1,-1,-1,-1)$. Assume we can imbed a 3D hypersurface $\Sigma_3$, imbedded in the 4D space $\Sigma_4$. Assume, in some place, we can define a vector $n$ normal to $\Sigma_3$, which satisfies:
\begin{equation*}
	n^0n_0+\left(n^1n_1+n^1n_1+n^1n_1\right)>0
\end{equation*}
which, in the familiar language of special relativity theory, implies that $\Sigma_3$ is ``oriented in space'' (whereas the vector $n$, normal to $s$, is ``oriented in time'').

We introduce in the surface $\Sigma_3$ three coordinates $x^{\ast 1}$, $x^{\ast 2}$, $x^{\ast 3}$ which serve to characterise the point $P^{\ast} \in \Sigma_3$. Through each point $P^{\ast}$ of the 3D surface $\Sigma_3$ we draw the geodesic which is orthogonal to $\Sigma_3$ at $P^{\ast}$. These geodesic will form a non-intersecting curves in some neighborhood $M$ of $\Sigma_3$ such that, through each point $P$ of $M$ there will be exactly one of the geodesics constructed. We introduce now, in the entire 4D domain $M$, coordinates as follows: Given $P$, we consider the geodesic passing through $P$ and its original point $P^{\ast} \in  \Sigma_3$. We define the coordinate $x^i$ of $P$ in terms of the arc length $P^{\ast}P$ of the geodesic and of the coordinate $x^{\ast i}$ of $P ^{\ast}$.

In this manner, the three coordinates $x^1$, $x^2$, $x^3$ remain constant along any geodesic perpendicular to $\Sigma_3$. It follows that, along such a geodesic,
\begin{equation*}
	\diff s^2=\left(\diff\dot{x}\right)^2, \quad g_{00}=1.
\end{equation*}
\end{quote}

This is the classical way Gaussian coordinates are defined. This is possible if and only if $g_{00} \neq 0$, if the term of the line element related to time-marker is nonzero. Note that, in Fig. 1, through various choices of coordinate systems, in Schwarzschild solution this term $g_{00}$ (called $g_{tt}$) vanishes on the so-called horizon ($r = R_{\mathrm{s}}$), or throat surface ($\rho = 0$). It means that on this peculiar portion of the hypersurface, normal vector and space orientation cannot be defined.

\begin{figure}
	\begin{center}
		\includegraphics[width=6cm]{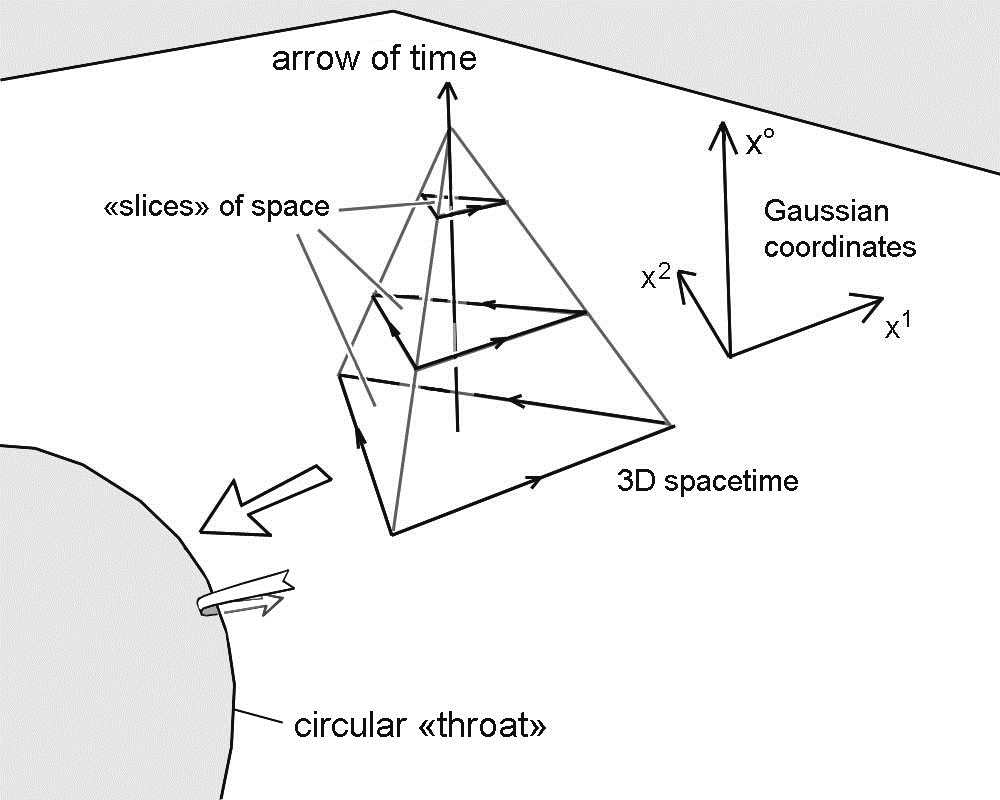}
	\end{center}
\caption{\label{fig:spacetime}Building a 3D spacetime.}
\end{figure}

Let us build a 3D spacetime on a 2D surface, as the twofold cover of a manifold $M_2$, as shown in \prettyref{fig:spacetime}.

\begin{figure}
	\begin{center}
		\includegraphics[width=6cm]{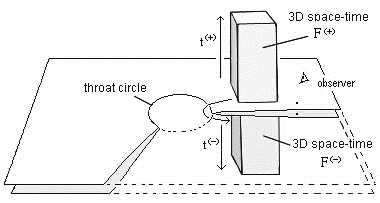}
	\end{center}
\caption{\label{fig:throat} In turning the throat circle, the arrow of time is inverted.}
\end{figure}

As shown in \prettyref{fig:throat} we will find some problem at the common boundary, where it is impossible to define space orientation and arrow of time (identified to normal vector). But, for adjacent regions we have imbricated PT-symmetrical spacetime structures. Such coupling concept, called at that time ``twin universe theory'', was first presented by Sakharov in 1967\cite{sakharov1967,sakharov1979,sakharov1980,sakharov1982} and later in Ref. \citenum{petit1977}. In addition, this goes with recent works\cite{petit2014a,petit2014b} (Janus Cosmological Model).

We have to deal with 4D spacetime, not 3D spacetime, which is just a didactic image of such geometric structure.

Back to the expression of the Schwarzschild metric in the new coordinates \eqref{eq16} we see that, on the throat surface ($\rho = 0$), $g_{tt}$ vanishes (as well as in expressions shown in \prettyref{fig:table}). According to Sec. 2.6 of Ref. \citenum{abs}, Gaussian coordinates cannot be defined, which means that on the throat surface, time and space cannot be oriented. This fits time and space inversion from fold $F ^{(+)}$ to fold $F^{(-)}$, so that we must write joint metrics:
\begin{subequations}
	\begin{multline}
		\diff s^{(+)2}=\frac{\Log\ch\rho}{1+\Log\ch\rho}c^2\diff t^{(+)2}\\-R_{\mathrm{s}}^2\left[\frac{\left(1+\Log\ch\rho\right)}{\Log\ch\rho}\tth^2\rho\diff\rho^2+\left(1+\Log\ch\rho\right)^2 \left(\diff\theta^2+\sin^2\theta\diff\varphi^2\right)\right],
	\end{multline}
	
	\begin{multline}
		\diff s^{(-)2}=\frac{\Log\ch\rho}{1+\Log\ch\rho}c^2\diff t^{(-)2}\\-R_{\mathrm{s}}^2\left[\frac{\left(1+\Log\ch\rho\right)}{\Log\ch\rho}\tth^2\rho\diff\rho^2+\left(1+\Log\ch\rho\right)^2 \left(\diff\theta^2+\sin^2\theta\diff\varphi^2\right)\right],
	\end{multline}
\end{subequations}
with
\begin{equation}
	t^{(-)}=-t^{(+)}.
\end{equation}

In the neighborhood of $\rho = 0$ it is possible to write the nearby expressions:
\begin{subequations}
	\begin{equation}
		\left(\diff s^{(+)}\right)^2=\frac{\rho^2}{2}\left(\diff t^{(+)}\right)^2-2\diff\rho^2-R_{\mathrm{s}}^2\diff\varphi^2,
	\end{equation}
	
	\begin{equation}
		\left(\diff s^{(-)}\right)^2=\frac{\rho^2}{2}\left(\diff t^{(-)}\right)^2-2\diff\rho^2-R_{\mathrm{s}}^2\diff\varphi^2.
	\end{equation}
\end{subequations}

The inversion of the time variable does not imply a change in the sign of the proper time, and from one fold to the other $\diff s^{(-)}$ takes the reins of $\diff s^{(+)}$. It is not possible to have an inversion in the length measure along a geodesic. In other words we have $\left(\diff s^{(+)}\right)\left(\diff s^{(-)}\right) > 0$. But $t^{(+)}$ and $t^{(-)}$ are nothing more than ``time markers'', simple coordinates, and thus one will have: $\diff t^{(-)} = -\diff t^{(+)}$. So here one finds again the central idea of differential geometry: the length element has an only intrinsic reality. Lagrange equations give always in the vicinity of $\rho = 0$ the relations:

\begin{equation}
	\ddot{\varphi}=\ddot{\rho}=0.
\end{equation}

These functions are linear and monotonic as a function of the proper times (lengths $s^{(+)}$ and $s^{(-)}$) that enchain themselves in passing from one fold to the other without inversion.

\begin{equation}
	\frac{\diff t^{(+)}}{\diff s^{(+)}}=\frac{C}{\rho^2}, \quad \frac{\diff t^{(-)}}{\diff s^{(-)}}=\frac{-C}{\rho^2}, \quad C=C\mathit{st}.
\end{equation}

The sign of the constant $C$ depends on the sense adopted for the passage from one fold to the other. The object is thus a ``black hole'' and a ``white fountain'' at the same time. But, if measured with variable $t^{(+)}$ the passage is achievable only in an infinite time. If the object results from the implosion of a neutron star, its mass would be transferred to the negative energy region. But for the observers located in one of the folds such phenomena of implosion--explosion will appear to be ``freezed in time''.

As shown in Ref. \citenum{souriau} time-inversion goes with mass and energy inversion. In Refs. \citenum{petit1977} and \citenum{sakharov1967}, according to Janus Cosmological Model the universe is composed by positive and negative energy (and mass if they own) particles, respectively described by metrics $g_{\mu\nu}^{(+)}$ and $g_{\mu\nu}^{(-)}$, solutions of a coupled field equation system. Spacetime bridges operate mass inversion. From Ref. \citenum{petit1977} we know that masses with opposite signs repel each other. On another hand negative matter does not interact with positive matter by electromagnetic, strong or weak interaction.

\section{Conclusion and Discussion}

If black holes exist and swallow matter this last, instead to be crushed in a central singularity, would be discretely rejected as an invisible negative matter.

The main feature of the theory presented here is the mass (and space) inversion process. The advantage is to avoid the puzzling problem of a ``central singularity'' and to explain the fate of matter swallowed by black holes. But it implies injection of negative energy (and mass if the own) particles in spacetime, considered as a manifold plus two metrics $g_{\mu\nu}^{(+)}$ and $g_{\mu\nu}^{(-)}$.

As precised in Ref. \citenum{petit2014a} we assumed that particles of opposite masses do not interact neither by electromagnetic forces nor strong or weak forces, they could not enter into a collision. Some colleagues have criticized this idea arguing that ``the particles are on the same spacetime''. The answer to this question is: if one considers the problem on purely geometrical grounds, those encounters would be ``geometrically impossible'' because the two subsets move along disjoint families of geodesics.

A second criticism may rely on the immediate instability of a quantum vacuum which could create pairs $(+m, -m)$. But it is based on the theoretical framework of quantum gravity that, however, still today remains purely hypothetical. The creation and annihilation of pairs of particles of opposite mass has not been de-scribed till today.

A third criticism may issue from Quantum Field Theory, which excludes straight away states of negative energy \emph{``because a particle could not have an energy less than that of the vacuum''} (p.~76 of Ref. \citenum{weinberg}). We quote:

\begin{quoting}
If we suppose that $\opT$ is linear and unitary then we should face the disastrous conclusion that for any state $\Psi$ of energy $E$ there is another state $\opT^{-1}\Psi$ of energy $-E$. To avoid this we are forced here to conclude that $\opT$ is anti-linear and anti-unitary.
\end{quoting}

In order to refute this statement we would say that it is bootstrap talk and that the conclusion is contained in the hypothesis, as occurs with the ``CPT theorem''.

In his book Weinberg,\cite{weinberg} let us quote his sentence on p. 104:

\begin{quoting}
No examples are known of particles that furnish unconventional representation of inversions, so these possibilities will not be pursued here. From now on, the inversions will be assumed to have the conventional action assumed in Sec. 2.6.
\end{quoting}

This sentence refers of course to the hypothesis expressed on p. 76 of Ref. \citenum{weinberg} about the anti-unitary and anti-linear character of the $\opT$ operator. However, cosmic acceleration implies the action of a negative pressure and hence of negative energy (pressure is an energy density by unit volume). The discovery of such quite un-foreseen phenomenon\cite{riess1998,perlmutter,riess2000,filippenko,leibundgut,knop,tonry,barris,riess2004,schmidt} makes it compelling for Quantum Field Theory to be extended in order to include negative energy states.

\section*{Erratum\label{erratum}}

Equation \eqref{eq13} page~\pageref{eq13} has a typo (extra square) and should be written:
\begin{equation*}
	r=R_{\mathrm{s}}+\frac{z^2}{4R_{\mathrm{s}}}
\end{equation*}

\end{document}